\documentstyle[preprint,aps,prabib]{revtex}
\begin{document}
\draft
\title{Quantization of strongly interacting fields}
\author{V. Dzhunushaliev
\thanks{E-mail address: dzhun@freenet.bishkek.su}}
\address{Dept. of Physics, Virginia Commonwealth University, 
Richmond, VA 23284-2000 and Theor. Physics Dept.,
Kyrgyz State National University,
720024, Bishkek, Kyrgyzstan}
\author{D. Singleton
\thanks{E-mail address : das3y@erwin.phys.virginia.edu}}
\address{Dept. of Physics, Virginia Commonwealth University,
Richmond, VA 23284-2000}
\date{\today}
\maketitle
\begin{abstract}
While perturbative techniques work extremely well for weakly interacting
field theories ({\it e.g.} QED), they are not useful when studying
strongly interacting field theories ({\it e.g.} QCD at low energies).
In this paper we review Heisenberg's idea about 
quantizing strongly interacting, non-linear fields,  
and an approximate method of solving the
infinite set of Tamm-Dankoff equations is suggested. We then 
apply this procedure to an infinite energy, classical flux 
tube-like solution of SU(2) Yang-Mills theory and show that this
quantization procedure ameliorates some of the bad behaviour of
the classical solution. We also discuss the possible application of this
quantization procedure to a recently proposed strongly interacting 
phonon model of High-$T_c$ superconductors.

\end{abstract}
\narrowtext

\section{Introduction}

In Ref. \cite{dzh1} a string model of High-$T_c$ superconductivity
was suggested. This model is based on the proposal that phonons have 
a strong self interaction. In this case a flux tube filled with
phonons appears to form between the Cooper electrons in a manner
analogous to QCD where it is often postulated that the strong interaction 
of the theory leads to the formation of confining flux tubes between
quarks. Just as in QCD where the formation of such 
flux tubes leads to a strong binding of the quarks up to a very 
high temperature when a deconfining phase sets in, so too in
the strongly interacting phonon picture the formation of such
flux tubes is conjectured to raise the temperature at which 
the Cooper pairing is broken. The possibility of such a strong
interaction between phonons in superconductors was also discussed 
in Ref. \cite{hrk}. Such strongly interacting theories can present
a challenge in that it is not possible to employ standard perturbation
theory techniques ({\it i.e.} Feynman diagrams) to them. Some time
ago Heisenberg conceived of the difficulties in using an expansion
in small parameters to strongly interacting quantum field theories
as a result  of his investigations into the Dirac equation with
nonlinear terms (the Heisenberg equation -- see, for example
Ref's \cite{h1} \cite{h2}). In these papers he repeatedly
underscored that a nonlinear theory with a strong coupling requires
the introduction of another quantization procedure. To this end
he worked out a quantization method for strong nonlinear fields
using the Tamm-Dankoff method. After briefly reviewing Heisenberg's
ideas we apply his quantization method to a classical flux tube-like
solution of the SU(2) Yang-Mills to show how the this quantization
procedure can soften some of the bad behaviour of the classical
solution.

\section{Heisenberg Quantization of Strongly Interacting Fields}

Heisenberg's basic idea proceeds from the fact that the n-point Green's
functions must be found from some infinite set of differential equations
derived from the field equations for the field operators. As an example
we present Heisenberg's method of quantization for a spinor field with
nonlinear self interaction.

The basic equation (Heisenberg equation) has the following form :
\begin{equation}
\label{1}
\gamma ^{\mu} \partial _{\mu} \psi (x) - l^2 \Im [\psi ({\bar \psi}
\psi) ] = 0
\end{equation}
where $\gamma ^{\mu}$ are Dirac matrices; $\psi (x), {\bar \psi}$ are 
the spinor field and its adjoint respectively; $\Im [\psi ({\bar \psi}
\psi) ] = \psi ({\bar \psi} \psi)$ or $\psi \gamma ^5 ({\bar \psi} \gamma ^5
\psi)$ or $\psi \gamma ^{\mu} ({\bar \psi} \gamma _{\mu} \psi)$ or
$\psi \gamma ^{\mu} \gamma ^5 ({\bar \psi} \gamma _{\mu} \gamma ^5
\psi )$. The constant $l$ has units of length, and sets the scale for the
strength of the interaction. Heisenberg 
emphasized that the 2-point Green's function,
$G_2 (x_2, x_1)$ in this theory differs strongly from the propagator in 
a linear theory. This difference lies in its behaviour on the light
come : in the nonlinear theory $G_2 (x_2 , x_1)$ oscillates strongly on
the light cone in contrast to the propagator of the linear theory
which has a $\delta$-like singularity. Heisenberg introduces the 
$\tau$ functions as
\begin{equation}
\label{2}
\tau (x_1 x_2 ... | y_1 y_2 ...) = \langle 0 | T[\psi (x_1) \psi (x_2) ...
\psi ^{\ast} (y_1) \psi ^{\ast} (y_2) ...] | \Phi \rangle
\end{equation}
where $T$ is the time ordering operator; $| \Phi \rangle$ is a state for
the system described by Eq. (\ref{1}). Eq. (\ref{2}) allows us to 
establish a one-to-one correspondence between the system state,
$| \Phi \rangle$, and the set of functions $\tau$. This state can be defined
using the infinite function set of Eq. (\ref{2}). Applying Heisenberg's
equation (\ref{1}) to (\ref{2}) we obtain the following infinite
system of equations for various $\tau $'s
\begin{eqnarray}
\label{3}
l^{-2} \gamma ^{\mu} _{(r)} {\partial \over \partial x^{\mu} _{(r)}}
&&\tau (x_1 ...x_n |y_1 ... y_n ) = \Im [ \tau (x_1 ... x_n x_r |
y_1 ... y_n y_r)] + \nonumber \\ 
&&\delta (x_r -y_1) \tau ( x_1 ... x_{r-1} x_{r+1} ... x_n |
y_2 ... y_{r-1} y_{r+1} ... y_n ) + \nonumber \\
&&\delta (x_r - y_2) \tau (x_1 ... x_{r-1} x_{r+1} ... x_n |
y_1 y_2 ... y_{r-1} y_{r+1} ... y_n ) + ...
\end{eqnarray}
Eq. (\ref{3}) represents one of an infinite set of coupled equations 
which relate various order (given by the index $n$) of the $\tau$ 
functions to one another. To make some head way toward solving 
this infinite set of equations Heisenberg employed the Tamm-Dankoff
method whereby he only considered $\tau$ functions up to a certain
order. This effectively turned the infinite set of coupled equations 
into a finite set of coupled equations.

The standard Feynman diagram technique of dealing with field 
theories via an expansion in terms of a small parameter does
not work for strongly, coupled nonlinear fields. Heisenberg 
used the procedure which is skecthed above to study the
Dirac equation with a nonlinear coupling. From a more recent 
perspective it may be interesting to apply the same procedure 
to nonlinear, bosonic field theories such as QCD in the low
energy limit or the recently proposed \cite{dzh1} strongly 
interacting phonon theory of High-$T_c$ superconductors. In
this paper we will apply the Heisenberg method to an  
infinite energy, flux tube-like solution for classical SU(2)
Yang-Mills theory. Under certain assumptions we find that
the unphysical behaviour of the classical SU(2)
solution is ``smoothed'' out when the Heisenberg technique
is applied. The formation of flux tubes is an important
feature of both QCD (a confining flux tube is thought
to form between two quarks) and the strongly interacting
phonon model (a flux tube is thought to form between the
Cooper electrons, binding them at higher temperatures that
is possible in the BCS picture).

\section{Quantization of SU(2) flux tube solution}

First we begin by discussing briefly the classical flux tube-like
solution to the SU(2) Yang-Mills theory. The sourceless Yang-Mills
equations for SU(2) are
\begin{equation}
\label{4}
\nabla _{\mu} F^a _{\mu \nu} = 0
\end{equation}
where $\nabla _{\mu} = \partial _{\mu} - i g A_{\mu} ^a T^a$ is the
covariant derviative; $T^a$ is an element of the group in some
representation; $F^a _{\mu \nu} = \partial _{\mu} A^a _{\nu}
- \partial _{\nu} A^a _{\mu} + g \epsilon^{abc} A_{\mu} ^b 
A_{\nu} ^c$ is the SU(2) field strength tensor; $A_{\mu} ^a$ is
the SU(2) gauge potential.

To simplify these Yang-Mills equations we make the following 
cylinderical symmetric ansatz
\begin{mathletters}
\label{5}
\begin{eqnarray}
A^1_t & = & f(\rho ),
\label{5:1}\\
A^2_z & = & v(\rho ),
\label{5:2}\\
A^3_{\varphi} & = & \rho w (\rho )
\label{5:3}
\end{eqnarray}
\end{mathletters}
here $z, \rho , \varphi$ are the standard cylinderical coordinates.
Substituting Eqs. (\ref{5}) into Eq. (\ref{4}) the Yang - Mills 
equations become
\begin{mathletters}
\label{6}
\begin{eqnarray}
f'' + \frac{f'}{\rho} & = & f\left (v^2 + w^2 \right ),
\label{6:1}\\
v'' + \frac{v'}{\rho} & = & v\left (-f^2 + w^2 \right ),
\label{6:2}\\
w'' + \frac{w'}{\rho}  - \frac{w}{\rho ^2}& = & w
\left (-f^2 + v^2 \right ),
\label{6:3}
\end{eqnarray}
\end{mathletters}
We further simplify these equations by taking $w =0$ which yields
\begin{mathletters}
\label{6a}
\begin{eqnarray}
f'' + \frac{f'}{\rho} & = & fv^2 ,
\label{6a:1}\\
v'' + \frac{v'}{\rho} & = & -vf^2 .
\label{6a:2}
\end{eqnarray}
\end{mathletters}
These equations can be solved numerically. When this is done it is
found that the ansatz function $f$ increases linearly while
$v$ is a strongly oscillating function \cite{dzh2}. 
The asymptotic behaviour of the ansatz functions $f, v$ confirms these
numerical calculations
\begin{mathletters}
\label{7}
\begin{eqnarray}
f & \approx & 2\left [x + \frac{\cos \left (2x^2 + 2\phi _1\right )}
{16x^3} \right ] ,
\label{7:1}\\
v & \approx & \sqrt{2} \frac{\sin \left (x^2 + \phi _1 \right )}{x} ,
\label{7:2}
\end{eqnarray}
\end{mathletters}
where $x= \rho / \rho _0$ is a dimensionless radius, and $\rho _0,
\phi _1$ are constants. The linearly increasing potential given
by the ansatz $f$ is very suggestive of the phenomenological
linear confining potentials of QCD. This classical solution
has a badly behaved field energy. The energy density for this
solution has the following asymptotic proportionality
\begin{equation}
\label{8}
{\cal E}  \propto  f'^2 + v'^2 + f^2v^2 \approx const,
\end{equation}
where Eqs. (\ref{7}) have been used.
Depending on the initial conditions of the solution
the energy density near $\rho = 0$ will be either a hollow ({\it i.e.}
an energy density less than the asymptotic value) or a hump ({\it i.e.}
an energy density greater than the asymptotic value). 
On account of this and the cylindrical symmetry of this solution 
we call this the ``string'' solution. The quotation marks 
indicate that this is a string from an energetic point of view, not from 
the potential ($A^a_{\mu}$) or field strength ($F^a_{\mu\nu}$) point of view.
The defect in this solution is made apparent when one calculates its
total field energy. To do this one must integrate the
energy density over all space. Eq. (\ref{8}) implies that this will
give an infinite answer. By applying the Heisenberg quantization
method to this system we find that this undesirable behaviour
of the classical solution is lessened. In order to simplify
Heisenberg's quantization method to the present nonlinear equations
we make the following assumptions :

1. The degrees of freedom relevant for studying this flux tube-like
solution (both classically and also quantum mechanically) are given
entirely by the two ansatz functions $f ,v$ of Eqs. (\ref{5}). No other
degrees of freedom arise through the quantization process.

2. From Eq. (\ref{7:1}) $f$ is a smoothly varying function for large
$x$, while $v$ is strongly ocsillating. Thus we take $f(\rho)$ to be
almost a classical degree of freedom while $v(\rho)$ is treated as a
fully quantum mechanical degree of freedom. Naively one might think that
in this way only the behaviour of $v$ would change while $f$ stayed
the same. However since $f$ and $v$ are interrelated due to the 
nonlinear nature of the equations of motion we find that both functions 
are modified.

To begin using Heisenberg's quantization scheme to this Yang-Mills
system we replace the ansatz functions by operators ${\hat f} (\rho) ,
{\hat v} (\rho )$.
\begin{mathletters}
\begin{eqnarray}
\label{9}
{\hat f}'' + {{\hat f}' \over x} &=& {\hat f} {\hat v}^2
\label{9:1} \\
{\hat v}'' + {{\hat v}' \over x} &=& -{\hat v} {\hat f}^2
\label{9:2}
\end{eqnarray}
\end{mathletters}
here the prime denotes a derivative with respect to $x$. Taking into
account assumption (2) we let ${\hat f} \rightarrow f$ become just
a classical function again, and replace ${\hat v}^2$ in Eq. ({\ref{9:1})
by its expectation value
\begin{mathletters}
\begin{eqnarray}
\label{10}
f'' + {f' \over x} &=& f \langle v^2 \rangle 
\label{10:1} \\
{\hat v}'' + {{\hat v}' \over x} &=& - {\hat v} f^2 
\label{10:2}
\end{eqnarray}
\end{mathletters}
Now if we took the expectation value of Eq. (\ref{10:2}) and
ignored the coupling to $f$ on the right hand side we would
have an equation for determining $\langle v \rangle = \langle 0 |
{\hat v} | 0 \rangle$. However the two nonlinear terms on the 
righthand side of Eqs. (\ref{10:1} - \ref{10:2}) show that a
new object, $\langle v^2 \rangle$ enters the picture so that
Eqs. (\ref{10:1} - \ref{10:2}) are not closed. To obtain an
equation for $\langle v^2 \rangle$ we act on ${\hat v} ^2 (x)$
with the operator 
$\left( {d^2 \over dx^2} + {1 \over x} {d \over dx} \right)$ giving
\begin{equation}
\label{11}
({\hat v}^2)'' + {1 \over x} ({\hat v}^2)' = -2 {\hat v}^2 f^2
+ 2 ({\hat v}')^2
\end{equation}
Taking the expectation value of this equation gives the desired
equation for $\langle v^2 \rangle$
\begin{equation}
\label{12}
\langle v^2 \rangle '' + {1 \over x} \langle v^2 \rangle ' =
-2 \langle v^2 \rangle f^2 + 2 \langle (v') ^2 \rangle
\end{equation}
Again this equation is not closed due to the $\langle (v') ^2 \rangle$ 
term. We could again try to find an equation for $\langle
(v') ^2 \rangle$ by the same procedure we employed for $\langle
v^2 \rangle$. This equation would also not be closed. Continuing
in this way we would find an infinite set of equations. In order
to have some hope of handling this problem we need to make some
approximation to cut this process off at some finite set of equations.
We try two different approximations for the $\langle (v') ^2 \rangle$
term and show that both yield similiar large $x$ behaviour that 
fixes the infinite field energy of the classical solution. First we
assume that $\langle (v') ^2 \rangle \approx \alpha \langle v^2 \rangle$
where $\alpha$ is some constant. This assumption yields the
following closed equation set
\begin{mathletters}
\begin{eqnarray}
\label{13}
\langle v^2 \rangle '' + {1 \over x} \langle v^2 \rangle '
&=& \langle v^2 \rangle (1 -f^2)
\label{13:1} \\
f'' + {1 \over x} f' &=& f \langle v^2 \rangle
\label{13:2}
\end{eqnarray}
\end{mathletters}
where we have rescaled the functions as : $\alpha ^2 x^2 \rightarrow
x^2, \langle v^2 \rangle / \alpha \rightarrow \langle v^2 \rangle ,
f/ \alpha \rightarrow f$. As $x \rightarrow \infty$ the asymptotic
form of the solution becomes
\begin{mathletters}
\begin{eqnarray}
\label{14}
\langle v^2 \rangle & \approx & v_0 ^2 {exp (-\gamma x) \over \sqrt{x}} 
\label{14:1} \\
f & \approx & f_{\infty} + f_0 {exp(-\gamma x) \over \sqrt{x}}
\label{14:2}
\end{eqnarray}
\end{mathletters}
with
\begin{equation}
\label{15}
f_0 = {f_{\infty} v_0 ^2 \over 2(1-f_{\infty} ^2)} , \; \; \; \; \; \; 
\gamma = \sqrt{2 (1-f_{\infty} ^2)}
\end{equation}
Thus if $|f_{\infty} | \le 1$ then $\gamma >0$ and we find that the 
quantum effects tend to modify the bad long distance behaviour of
both ansatz functions. 

Instead of using the assumption $\langle (v') ^2 \rangle \approx \alpha 
\langle v^2 \rangle$ to close the equations we could also have made
the assumption that $\langle (v') ^2 \rangle \approx \pm \langle v^2
\rangle '$. Since $\langle (v') ^2 \rangle$ is positive definite one 
picks the $\pm$ sign so that the righthand side of this assumption
is also positive definite. Under this assumption the equations
become
\begin{mathletters}
\begin{eqnarray}
\label{16}
\langle v^2 \rangle '' +\left( {1 \over x} \mp 2 \right) \langle
v^2 \rangle ' &=& - 2 \langle v^2 \rangle f^2 
\label{16:1} \\
f'' + {1 \over x} f' &=& f \langle v^2 \rangle
\label{16:2}
\end{eqnarray}
\end{mathletters}
The approximate solution of Eqs. (\ref{16:1} - \ref{16:2}) again has the
same functional form as Eqs. (\ref{14:1} - \ref{14:2}) but now
\begin{equation}
\label{17}
f_0 \gamma ^2 = f_{\infty} v_0 ^2 \; \; \; \; \; \; \;
\gamma ^2 \pm 2 \gamma = - 2 f_{\infty} ^2
\end{equation}
The second relationship can be written (using the first relationship)
as
\begin{equation}
\label{18}
\gamma = \mp f_{\infty} \left( f_{\infty} + {v_0 ^2 \over 2 f_0} \right)
\end{equation}
Although $\langle v'^2 \rangle \approx + \langle v^2 \rangle '$
leads to unphysical exponentially growing solutions, the
assumption $\langle v'^2 \rangle \approx - \langle v^2 \rangle '$ leads
to exponentially decaying solutions. Under this latter assumption 
and the previous assumption ($\langle v'^2 \rangle \approx
\alpha \langle v^2 \rangle$)  for cutting
off the equations we find that that the quantum mechanical treatment
of this nonlinear system modifies the bad features of the classical
solution. The asymptotic behaviour of $v$ goes from being strongly
ocsillating (see Eq. (\ref{7:2})) to decaying exponentially, while the
asymptotic behaviour of $f$ goes from being linearly increasing 
(see Eq. (\ref{7:1})) to also decaying exponentially. If the asymptotic
form for these ansatz functions are used in the energy density, ${\cal E}$,
of Eq. (\ref{8}) we find that the field energy is now finite. (To
calculate ${\cal E}$ we would replace the classcial terms $v'^2 , f^2 v^2$ 
by the appropriate quantum operator and take the expectation value.
The $\langle v'^2 \rangle$ would be handled according to the assumption
we used for closing the equations).

\section{Discussion}
In this work we have applied Heisenberg's ideas about quantizing strongly
interacting, nonlinear fields, not to fermionic fields as Heisenberg
did, but to the nonlinear field equations of an
SU(2) Yang-Mills gauge theory. Although the solution (Eqs. (\ref{7})) 
for the classical equations of motion (Eqs. (\ref{6a})) had some 
interesting features (the linear, confining behaviour indicated by
the ansatz function $f$ and the flux tube-like structure of the
energy density) these features also give the classical solution
the unphysical feature of having an infinite field energy. Using
Heisenberg's quantization procedure on this system and making some
assumptions in order to cut off the infinite equation set, we found 
that the quantum effects replaced the bad large distance classical
behaviour of $f,v$ with physically reasonable exponentially decaying
behaviour. This results in the field energy of the solution being
finite. Now in the small $x$ region one can still expect to find
the interesting behaviour ({\it i.e.} the linear increase of $f$)
of the classical solution due to asymptotic freedom \cite{gross}.
In non-Abelian theories the coupling strength can become small at 
small distance scales ({\it i.e.} small $x$) so that the classical solution
should be increasingly valid as $x \rightarrow 0$. Actually in this
$x \rightarrow 0$ limit one cannot use the asymptotic form of $f, v$
of Eq. (\ref{7}), but one must investigate the classical equations
numerically. When this is done \cite{dzh2} one again finds that
$f$ is approximately linearly increasing even as $x \rightarrow 0$.
The conclusion is that at short distances one has the interesting
features of the classical solution, while at large distances the
quantum effects, as taken into account via Heisenberg's method,
replace the bad long distance behaviour of the classical solution
with a physically reasonable behaviour.

This quantization procedure that we have applied to the SU(2)
flux tube may also be useful in investigating similiar structures
in other theories. In QCD it is suggested that the formation of flux 
tubes between isolated quarks is the mechanism responsible for 
confinement. However, due to the large coupling strength of QCD
the standard, perturbative treatment of the quantum effects of
the theory via Feynman diagrams is ruled out. The present method
may be useful for investigating the flux tube structures of QCD.
Recently, a nonlinear, strongly interacting phonon model of
High-$T_c$ superconductors was given \cite{dzh1}, where the
strong interaction between the phonons was postulated to lead
to the formation of a flux tube between the Cooper electrons.
This allowed the electrons to remain correlated up to much
higher temperatures that is normally found in the BCS model.
The quantization technique used here on the SU(2) Yang-Mills
theory may be useful in studying this model.

\section{Acknowledgements} This work has been funded in part by the
National Research Council under the Collaboration in Basic Science and
Engineering Program. The mention of trade names or commercial
products does not imply endorsement by the NRC. VD is grateful to
Prof. V.T. Gurovich for helpful discussions related to this topic.

\end{document}